# Realistic Multimedia Tools based on Physical Models: II. The Binary 3D Renderer (B3dR)

Ioannis Pachoulakis

Department of Applied Informatics & Multimedia, TEI of Crete
and
Centre for Technological Research of Crete
ip@epp.teicrete.gr

## Abstract

The present article reports on the second tool of a custom-built toolkit intended to train astronomers into simulating and visualizing the composite 3D structure of winds from hot close double stars by implementing a technique which is similar to multi-directional medical tomography. The flagships of the toolkit are the Spectrum Analyzer and Animator ($SA^2$) and the Binary 3D Renderer (B3dR). Following application of the first tool, $SA^2$ as detailed in paper I, the composite wind structure of the binary has been recovered and the B3dR is subsequently employed to visualize the results and simulate the revolution of the entire system (stars, winds and wind-interaction effects) around the common centre of mass. The B3dR thus repackages the end product of a lengthy physical modeling process to generate realistic multimedia content and enable the presentation of the 3D system from the point of view of an observer on Earth as well as from any other observer location in the Galaxy.

## Keywords

Realistic Multimedia Content, Physical Models, Interactivity, 3D visualization.

## 1. Introduction

Using UV spectra secured by the International Ultraviolet Explorer (IUE) satellite at various orbital phases, the present sequence of two articles (of which this is article-II), shows how to model the wind spectral lines of hot close binaries, reconstruct the 3D composite wind structure and simulate the motion of the system around a common center of mass. The present article details the application of the tool Binary 3D Renderer (B3dR). The output is realistic content that can be incorporated in multimedia applications and aims at increasing the effectiveness and communication impact of research results.

## 2. The Binary 3D Renderer (B3dR)

As discussed in paper-I (Pachoulakis, 2008), the fit quality for the synthetic binary wind-line profiles has been conditioned solely by the requirement that the observed wind profiles be reproduced as faithfully as possible. A way to quantify the quality of the fit is to study the phase-locked patterns of the "model minus observed" - in the sense [BSEI-Observed] - residuals, a process that is automated with the help of the $SA^2$ as follows:
- First, each (normalized) BSEI profile is multiplied by the same straight-line continuum that was used earlier to normalize the corresponding observed wind-line profile, a process that yields a sequence of model profiles in units of absolute flux.

- Since BSEI does not model line blanketing, each model profile is also multiplied by $1/(1-z)$, where $z$ is the level of line blanketing calculated from the tables of Kurucz (1979) for the appropriate stellar temperature and luminosity in the neighborhood of the wind profile being modeled. As a result, the line-blanketed BSEI model profiles are always brighter than the observed ones.
- Finally, the unmodeled residuals are integrated across the profile and the resulting measures are divided by the profile width to yield average monochromatic flux measures.

Following this recipe, the unmodeled residuals for HD159176 have been computed and modeled (Pfeiffer, Pachoulakis, Koch and Stickland, 1997) using the Binary Wind Interaction (BWI) code developed by R. J. Pfeiffer. BWI uses numerical grids to represent the stellar photospheres, the winds, and an (optional) wind inhomogeneity to partition the residuals into three fundamental categories of wind-interaction effects:
- Eclipses of portions of the remote photosphere, wind, or wind inhomogeneity by the nearer photosphere,
- Attenuations of portions of the remote photosphere, wind, or wind inhomogeneity by the nearer wind and,
- Attenuations of portions of the photospheres and winds of both stars by the wind inhomogeneity.

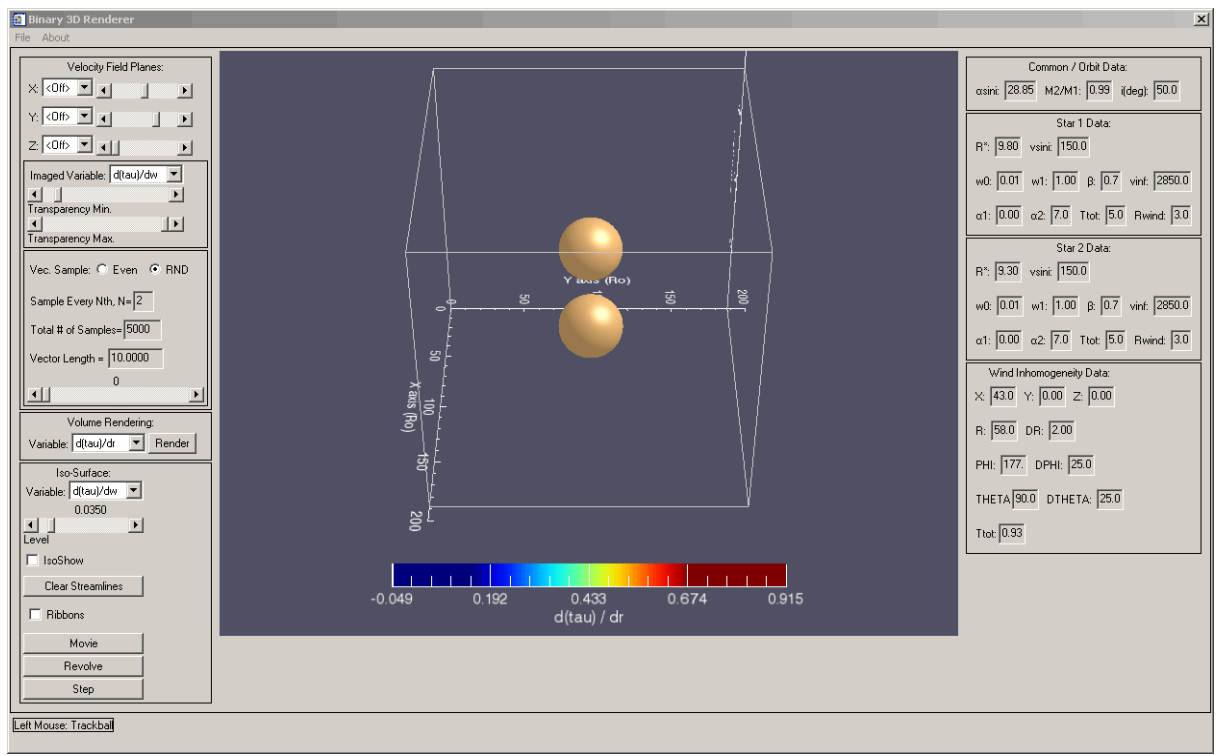

**Figure 1: Output of SA2 showing the BSEI fits to the observed NV and CIV profiles.**

To aid the visualization of the results, the Binary 3D Renderer (B3dR) has been devised to reconstruct a physically realistic simulation of a binary system which includes the stars, the winds and the wind-interaction effects. Figure 1 shows the organization of the user interface in three vertical panels. The middle (3D graphics) panel renders HD159176 at continuum wavelengths, where the winds and their interaction regions are invisible, so and the observer sees just the two photospheres. The rightmost panel is used to enter stellar, orbital and wind data published in radial velocity and wind analyses which the B3dR uses to compute and render any physical quantity of interest in the graphics panel.

In addition, with the help of the user controls on the leftmost panel, the user may isolate and focus on subsets of data (e.g., slices). Figure 2 shows some of the possibilities: the generation of vector plots for the velocity field at any given location (panels a and b), the production of images from data cross-sections and their annotation with ribbons showing magnitude and direction for the local velocity vector (panel c), as well as iso-surfaces of volume variables such as wind speed, optical depth, etc (panel d).

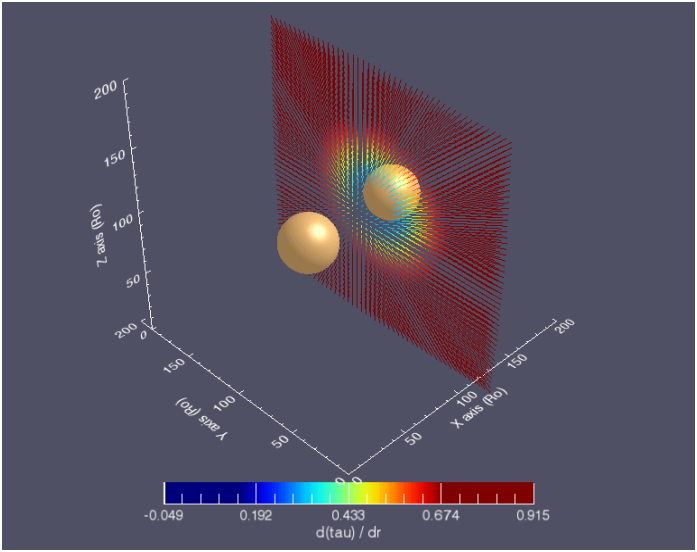

**Figure 2a: Velocity field plane normal to the line of star centers with regular vector sampling**

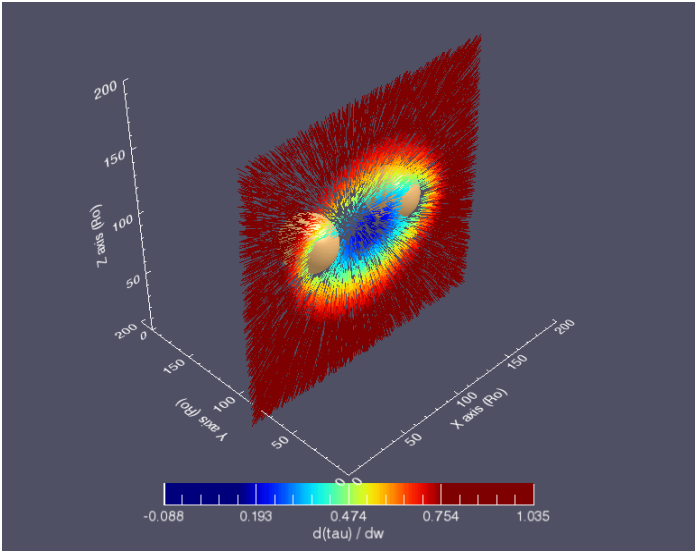

**Figure 2b: Velocity field plane normal to the Y-axis with random vector sampling**

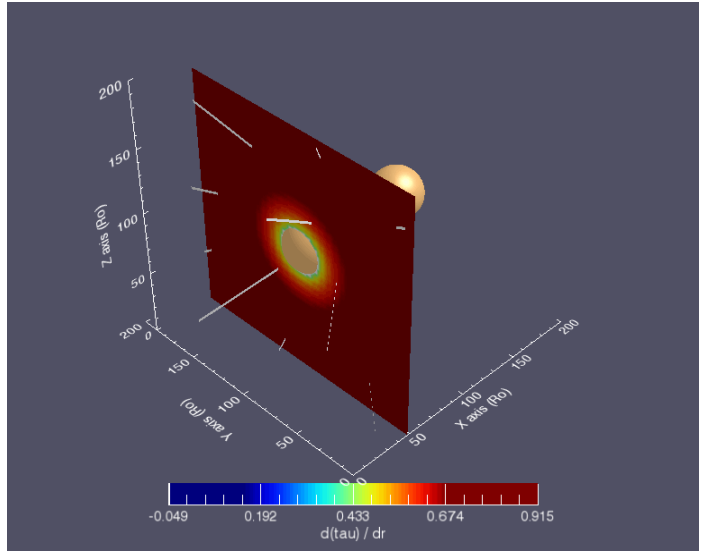

**Figure 2c: Color-coded velocity field plane with user-placed velocity ribbons**

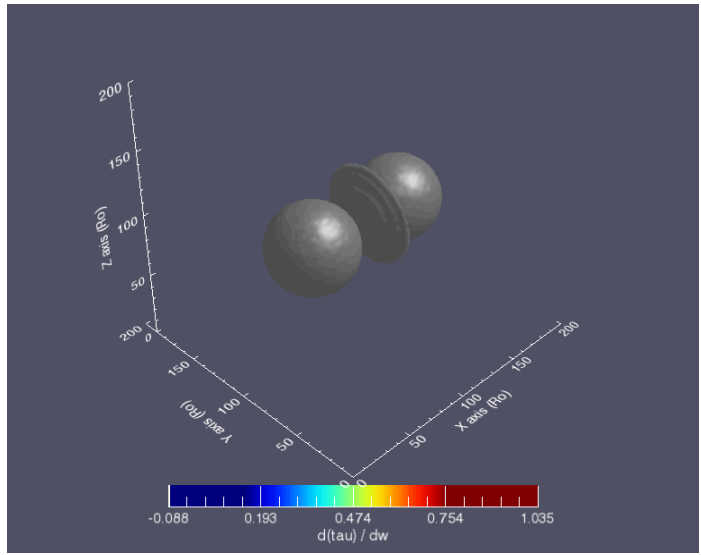

**Figure 2d: An iso-surface of optical depth**

The user may also volume-render physically interesting quantities and produce a simulation of what an observer near the binary system would "see" at wind-line wavelengths. Accordingly, Figure 3 shows a sequence of 3D renderings for HD159176 at CIV wind-line wavelengths and at an inclination of $50^o$ in the course of a single orbital cycle in steps of $\Delta\varphi=0.125$, starting from $\varphi=0.0$ in the top left panel. It is exactly these wind-line photons that propagate through these varying geometries in the direction to the observer that are modeled to unlock the structure of the composite wind. In fact, at an inclination $i=0^o$ the wind-line profiles of a hot binary would not show any phase-dependent variability at all and the method described in the preceding sections would be quite inapplicable.

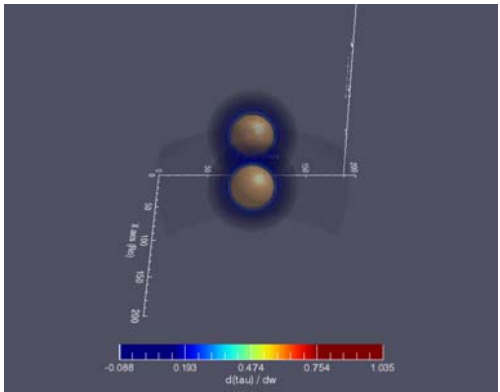
(a) φ = 0.000

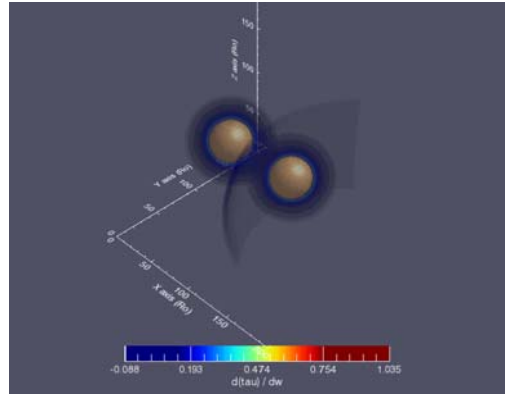
(b) φ = 0.125

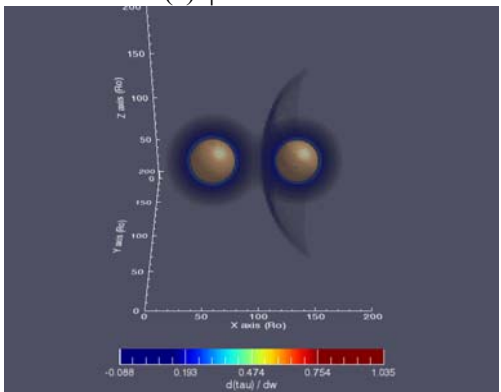
(c) φ = 0.250

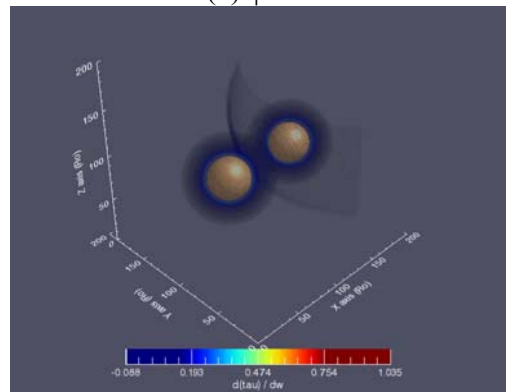
(d) φ = 0.375

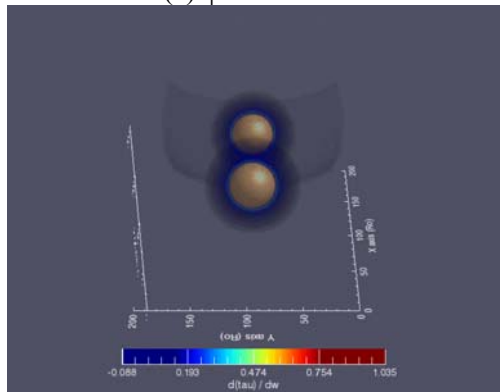
(e) φ = 0.500

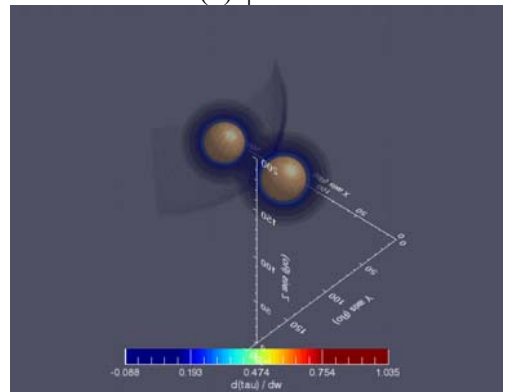
(f) φ = 0.625

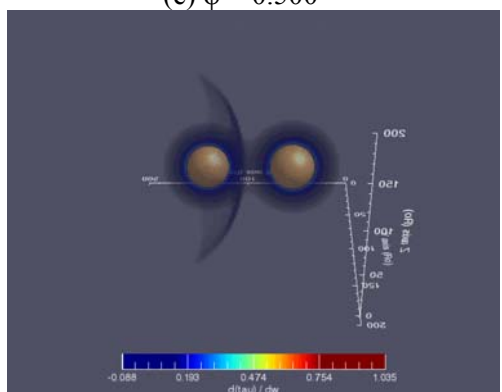
(g) φ = 0.750

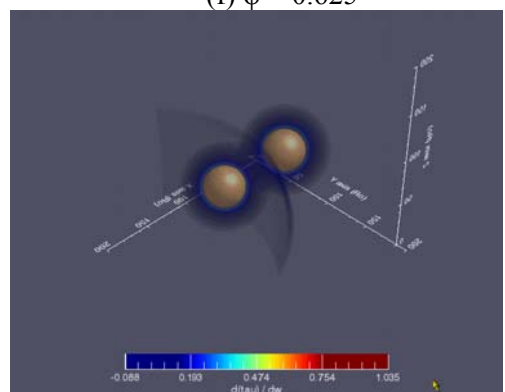
(h) φ = 0.875

**Figure 3:** The photospheres, composite wind and wind inhomogeneity at wind-line wavelengths of HD159176, using in B3dR. Panels (a) through (h) show the progress of an orbital cycle in steps of Δφ=0.125, starting from φ=0.0 at the top left.
Note: the boxy frame of Figure 2 has been removed and only the three axes appear to avoid screen clutter.

## 3. Discussion

A sequence of articles, of which the present is the second, reports on a method used to model, simulate and visualize the 3D structure of astrophysical wind volumes, which operates in a way similar to multi-directional medical tomography in that the spatial structure of an extended target can be reconstructed from a number of spectra obtained by scanning that target from several directions. The present article reports on the use of the tool Binary 3D Renderer (B3dR) to render the stellar photospheres and the composite binary wind, including the possible wind enhancement(s). Using published orbital data and absolute geometries, this model can be stepped in orbital phase to produce a visualization of what a nearby observer would actually "see" through a pair of spectacles operating at wind-line wavelengths as the stars revolve around their common center of mass.

In addition, the B3dR can be used to visualize the binary winds at any different orbital inclination. For example, at a different orbital inclination the non-eclipsing HD159176 could be made to eclipse and the same modeled binary wind solution would then yield a set of model wind line profiles with amplified phase-locked variability. In fact, among the spectroscopic (Doppler-resolved) OB binary population, those binaries that actually show photospheric eclipses (eclipsing binaries) are the most rewarding, as fundamental stellar parameters such as masses and radii can be accurately recovered from radial velocity and light curve solutions. It is then for such binaries, whose absolute orbital and stellar geometries are well established, that wind structure and interaction parameters can be tightly correlated to fundamental stellar parameters.

Although the present mode of modeling may be unable to recover weak shock fronts, the winds of HD159176 are sufficiently energetic to give rise to an interface shock whose signature is attested by the observed X-Ray surplus from the binary (DeBecker, Rauw, Pittard, Antokhin, Stevens, Gosset and Owocki, 2004; Oskinova, 2005}). In fact, because the stars are not static but revolve in their Keplerian orbits, such a layer does not have to be (and indeed is not) perpendicular to the stellar line of centers.

In summary, the modeling approach discussed in paper I (Pachoulakis, 2008) and the present article provides the grounds for a modeling framework for the understanding of the significantly more extravagant interactions which are of interest in hot double stars, offers a meaningful and consistent method that results in a homogeneous physical parameterization of the stellar winds and provides interactive tools (SA$^2$) to automate the analysis of the spectra and visualization tools (B3dR) which render and explore physically interesting quantities in 3D space.